# Institutional Diplomacy ////////////////////////

## A Glimpse of International Cooperation in Astrophysical Sciences in India

**Ram Sagar**
Indian Institute of Astrophysics, Sarajapur Road, Bangalore, India
Aryabhatta Research Institute of Observational Sciences (ARIES), Manora Peak, Nainital, India
*E-mail: ram_sagar0@yahoo.co.in; ramsagar@iiap.res.in*

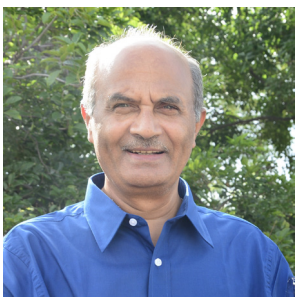

Astronomy and Astrophysics is an observational science dealing with celestial objects. Aryabhatta Research Institute of Observational Sciences (ARIES) is one of the premier institutions in astronomy and astrophysics and has contributed significantly to strengthening India's position in this field. No doubt, India is a part of several mega-science projects in the domain of Astronomy and Astrophysics, such as the Thirty Meter Telescope (TMT); Square Kilometer Array (SKA) and Laser Interferometer Gravitational-wave Observatory (LIGO) projects. India's growing engagement with mega-science projects has brought a positive impact on its science and technology (S&T) landscape.[1] A few such collaborations are briefed below.

**The Nainital 79/51 cm Schmidt Baker-Nunn (BN) camera**

From July 1957 to December 1958, a collaborative scientific program, the International Geophysical Year (IGY), was conducted to study the earth and its environment. IGY was sponsored by the International Council of Scientific Unions and involved nearly 30,000 scientists from over 70 countries, including India. It led to several significant scientific findings, including the discovery of the Van Allen radiation belts, defining of mid-ocean submarine ridges, the exploration of outer space, the construction of earth satellites, and increased research in the Arctic and Antarctic polar regions. IGY demonstrated the ability of all nations to work together harmoniously for the welfare of human endeavour. India contributed greatly to several IGY scientific activities, including optical photography of the artificial Earth satellites against the starry sky background. For this, the Smithsonian Astrophysical Observatory, United States of America (USA) installed twelve 79/51 cm Schmidt BN cameras along with a precision timing system[2] as part of a global network located in Argentina, Australia, Curacao, Iran, India (Nainital), Japan, Peru, Spain, South Africa and USA (New Mexico, Florida, Hawaii). Over 45,700 satellite transits, including those of India's Aryabhatta and Apollo-11, 12 and 17 of the USA, were successfully recorded from Nainital.[3]

**The Udaipur Solar Observatory and Global Oscillation Network Group (GONG)**

The GONG is an international project for a detailed study of solar internal structure and dynamics using helioseismology. Therefore, a network of six extremely sensitive and stable solar velocity imagers was installed at the Udaipur Solar Observatory in India; the Big Bear Solar Observatory in California, USA; the Mauna Loa Observatory in Hawaii, USA; the Learmonth Solar Observatory in Western Australia; the Observatorio del Teide in the Canary Islands and the Cerro Tololo Inter-American Observatory in Chile. Their longitudinal locations allow the GONG network to make 24/7 continuous observations of the Sun's "five-minute" oscillations. GONG project has been operational since October 1995 with the participation of about 175 individual members from over 20 nations, including India[4] and being implemented and coordinated by the National Solar Observatory's Integrated Synoptic Program, USA. Such international collaborations enable long-duration continuous observations that immensely help in examining the sun's variability.[5]

**Indo-Belgian collaboration**

ARIES hosts four telescopes of apertures 1.04 m, 1.3 m, 3.6 m and 4 m. The latter two are 'Indo-Belgian' telescopes. The Devasthal site (Longitude = 79°41'04" E, Latitude = 29°21'40" N and Altitude =2424±4 m), identified as a potential astronomical site for locating modern optical telescopes, became an Observatory in the year 2010 when 1.3 m Devasthal fast optical telescope (DFOT) started observing celestial objects. The 3.6 m Devasthal Optical Telescope (DOT) was successfully installed in 2015 and technically activated on 30 March 2016, jointly by the premiers of both countries, India and Belgium.[3] In April 2022, a 4 m International Liquid Mirror Telescope (ILMT) was established for the first time in India to observe the celestial bodies. This new telescope is designed and built by an international consortium from Belgium, Canada, India, Poland and Uzbekistan.[6] An aerial view of the Devasthal Observatory is shown in Figure 1.

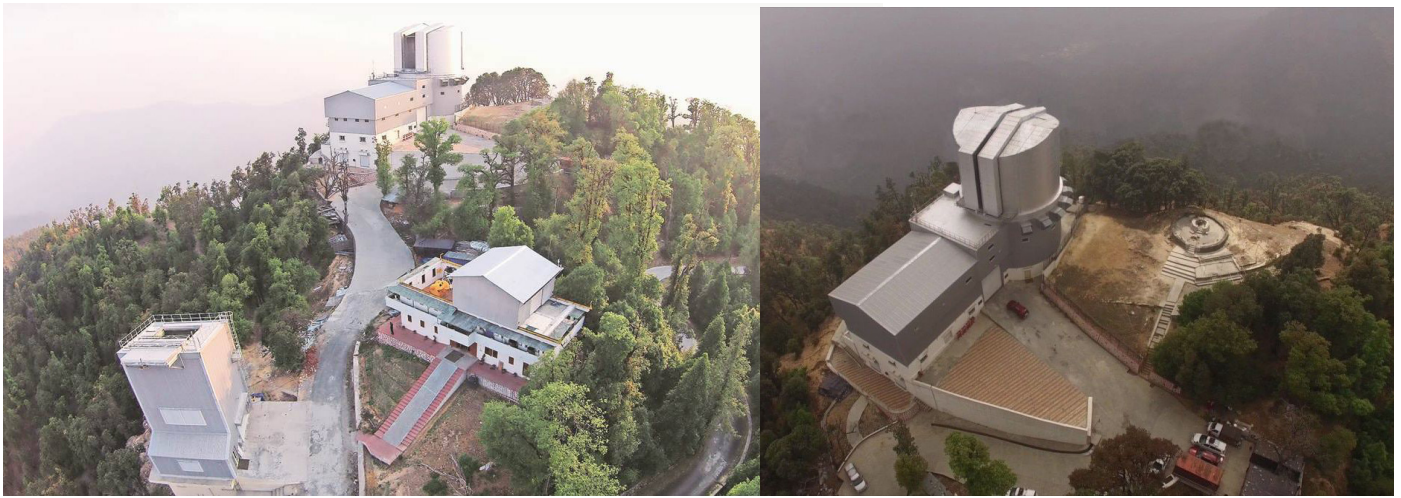

Figure 1. Devasthal Observatory hosting the (a) 1.3 m DFOT building with the sliding roof (bottom right) and rectangular building of the 4 m ILMT (bottom left), and (b) 3.6 m DOT near the adjacent Devasthal temple.

The Belgo-Indian Network for Astronomy and Astrophysics (BINA) was created to increase academic interactions between Indian and Belgian astronomers and engineers so that modern observing instruments for Devasthal telescopes can be developed jointly.[7] The third BINA workshop was held recently from 22 to 24 March 2023 in India. Several young and established researchers from both countries have benefitted from this collaboration. Overall, it has been an excellent example of diplomacy in science.

**Ongoing ground-based mega observational facilities and challenges**

The Indian astronomical community is participating in many international mega-science projects.[1] For instance, the TMT Observatory is being built by an international consortium of five countries, namely Canada, China, India, Japan and USA. TMT will be one of the world's advanced ground-based optical, NIR, and mid-infrared observatories since it – using cutting-edge global technology – will integrate the latest innovations in precision control, segmented mirror design, and adaptive optics.[8] Another mega project in this field is SKA - earth's largest next-generation radio telescope project. This project is funded by an international collaboration of more than a dozen countries, including India. The design of the SKA has recently been completed with the active participation of Indians. It has now entered the construction phase.[9]

The LIGO-India is another mega-science project. Once operational, it will establish a state-of-the-art advanced LIGO gravitational wave observatory on Indian soil in collaboration with the LIGO laboratory in the USA.[10] All these ground-based observational facilities are unique and complementary to each other and are essential for making Indian astrophysical science globally competitive. However, these mega-projects are marred by various challenges such as large size (physical), long timelines, massive costs, credit sharing issues, and cost and time uncertainties.[1]

**Space-based Indian Observatory**

India's first multiwavelength astronomy satellite, AstroSat, was successfully launched on 28 September 2015. In this space mission, national efforts were supplemented with international contributions from United Kingdom and Canada. It provided new bridges for global collaboration in the field of space astronomy. A few upcoming Indian space projects are summarised by Sreekumar and Rao,[11] including the Indian spectroscopic and imaging space telescope (INSIST)[12] and Daksha, a broadband high-energy all-sky mission.[13] All these Indian space missions are a product of international collaborations and are good illustrators of using diplomacy in science.

**Impact of international collaborations in the area of Astrophysical sciences**

From Indian Observatories, mostly celestial objects in the northern hemisphere are seen, while for proper understanding of the Universe, observations of both northern and southern hemisphere objects are essential. To overcome this, various global collaborations led by the author studied northern star clusters using observational facilities of Indian observatories, while southern star clusters and of Magellanic Clouds were studied using observations taken from Australia and Chile.[14-16]

Another example is the study of rapidly oscillating A-peculiar (ROAp) stars. Out of the 31 ROAp stars known till 1997, only three were discovered in the northern hemisphere. To investigate this big difference in the known ROAp stars in the two hemispheres, a survey programme named Nainital–Cape Survey was initiated in 1997 at Nainital. Under this international collaboration, northern and southern hemisphere objects were observed from Nainital and Cape Town, respectively. A few years of observations discovered 4 δ Sct pulsating variables and a ROAp northern hemisphere star.[3]

Indian astronomical observatories (Longitude ~ 79° E) are located in the middle of ~ 180° longitudinal gap between the Canary Islands (~20° W) and Eastern Australia (~160° E), having modern astronomical observational facilities. Consequently, they play a crucial role internationally as the observations which are not possible from both these countries due to daylight or other reasons can be taken from the Indian observatories or vice-versa, as demonstrated above. Such Indian observations have contributed to many discoveries, including time-domain astronomy, e.g., studies of Gamma-Ray Burst (GRB) afterglows and variable sources which require 24 h continuous monitoring. The extended ring system around Uranus was discovered with a combination of quality observations taken from Kavalur and Nainital in India; Cape Town in South Africa; Tokyo in Japan, and Western Australia during the occultation event of SAO 158687 by Uranus on 10 March 1977. In the same way, under international collaborations, Indian multiwavelength observations of GRB afterglows support the core collapse model for the progenitor of long-duration GRBs and put observational constraints on the popular progenitor's models of short-duration GRBs.[3,17] All these illustrate the importance of international collaborations in Astronomy and Astrophysics.

In the future, India may take the initiative to collaborate with neighbouring countries like Afghanistan, China, Hong Kong, Nepal, Pakistan, Singapore, Thailand and Uzbekistan, etc., to utilise their available expertise and resources optimally.

## Acknowledgements

Thanks to Shri S.K. Varshney, Head, International Division, Department of Science and Technology, New Delhi, for the invitation to write this article; the National Academy of Sciences, India, Prayagraj for the award of Honorary Scientist position and Director, IIA for hosting during this work.

## References

1. Asthana P (2022) Mega Science Programme in India – evolution and prospects. Current Science, 123(1): 26–36. https://doi.org/10.18520/cs/v123/i1/26-36
2. Bappu MKV (1958) Optical tracking of artificial satellites. Journal of Scientific & Industrial Research, 17A: 95–98.
3. Sagar R (2022) History of ARIES: a premier research institute in the area of observational sciences. Indian Journal of History of Science, 57(1): 227–247. https://doi.org/10.1007/s43539-022-00054-0
4. Leibacher JW (1999) The global oscillation network group (GONG) project. Advances in Space Research, 24(2): 173–176. https://doi.org/10.1016/S0273-1177(99)00497-4
5. Jain K, Tripathy SC, Hill F, Pevtsov AA (2021) Continuous solar observations from the ground-assessing duty cycle from GONG observations. Publications of the Astronomical Society of the Pacific, 133: 105001. https://doi.org/10.1088/1538-3873/ac24d5
6. Surdej J, Hickson P, Kumar B, Misra K (2022) First Light with the 4-m International Liquid Mirror Telescope. Physics News, 52(3): 25–28.
7. Joshi S, De Cat P (2019) Overview of the BINA activities. Bulletin de la Societe Royale des Sciences de Liege, 88: 19–30. https://doi.org/10.25518/0037-9565.8625
8. Reddy BE, Ramaprakash AN (2017) India's participation in the Thirty Meter Telescope international observatory project. Current Science, 113: 631–638. https://doi.org/10.18520/cs/v113/i04/631-638
9. Gupta Y (2022) The Square Kilometre Array: current status and India's role in this upcoming facility. Journal of Astrophysics and Astronomy, 43: 68. https://doi.org/10.1007/s12036-022-09856-1
10. Souradeep T, Raja S, Khan Z, Unnikrishnan CS, Iyer B (2017) LIGO-India–a unique adventure in Indian science. Current Science, 113(4): 672–677. https://doi.org/10.18520/cs/v113/i04/672-677
11. Sreekumar P, Rao VK (2021) Beyond AstroSat: Astronomy missions under review. Journal of Astrophysics and Astronomy, 42: 78. https://doi.org/10.1007/s12036-021-09744-0
12. Subramaniam A (2022) An overview of the proposed Indian spectroscopic and imaging space telescope, Journal of Astrophysics and Astronomy, 43: 80. https://doi.org/10.1007/s12036-022-09870-3
13. Bhalerao V et al. (2022) Daksha: On Alert for High Energy Transients. eprint arXiv:2211.12055. https://doi.org/10.48550/arxiv.2211.12055
14. Sagar R, Munari U, de Boer KS (2001) A multicolour CCD photometric and mass function study of the distant southern open star clusters NGC 3105, NGC 3603, Melotte 105, Hogg 15, NGC 4815, Pismis 20 and NGC 6253. Monthly Notices of the Royal Astronomical Society, 327: 23–45. https://doi.org/10.1046/j.1365-8711.2001.04438.x
15. Sagar R, Richtler T (1991) Mass functions of five Large Magellanic Cloud star clusters. Astronomy and Astrophysics, 250: 324–339.
16. Subramaniam A, Sagar R (1995) Young LMC star clusters as a test for stellar evolutionary models. Astronomy and Astrophysics, 297: 695–706.
17. Troja E et al. (2022) A nearby long gamma-ray burst from a merger of compact objects. Nature, 612: 228–231. https://doi.org/10.1038/s41586-022-05327-3